\documentclass{IEEEtran}
\usepackage{cite}
\usepackage{amsmath,amssymb,amsfonts}
\usepackage{graphicx}
\usepackage{textcomp,nicefrac}
\usepackage[switch, pagewise]{lineno} 
\usepackage{xcolor,soul,framed} 
\usepackage{siunitx}

\def\BibTeX{{\rm B\kern-.05em{\sc i\kern-.025em b}\kern-.08em
T\kern-.1667em\lower.7ex\hbox{E}\kern-.125emX}}
\begin{document}
\title{Combined Effects of Transient Ionizing and Electromagnetic Pulse on Vertical NPN Bipolar Transistor}
\author{Meiqing~Zhong,~\IEEEmembership{Student Member,~IEEE,}
      ~Cui~Meng,~\IEEEmembership{Senior Member,~IEEE,}
      ~Yinong~Liu,~\IEEEmembership{Member,~IEEE,}\\
      Lanfeng~Yuan, 
      Chicheng~Liu, 
      Bolun~Feng,
      and~Maoxing~Zhang,~\IEEEmembership{Member,~IEEE}
\thanks{Manuscript received December 28, 2024.}
\thanks{Meiqing Zhong, Yinong Liu, Lanfeng Yuan, Chicheng Liu and Bolun Feng are with the Department of Engineering Physics, Tsinghua University, Beijing 100084, China, and also with the Key Laboratory of Particle and Radiation Imaging, Ministry of Education, Beijing 100084, China (e-mail: zhongmq20@mails.tsinghua.edu.cn; liuyinong@mail.tsinghua.edu.cn).}
\thanks{Cui Meng and Maoxing Zhang are with the College of Electrical Engineering, Zhejiang University, Hangzhou 310027, China (e-mail: mengcui@zju.edu.cn).}}

\maketitle

\begin{abstract}
Combined effects of transient ionizing and electromagnetic pulse on vertical NPN bipolar transistor were experimentally investigated under pulsed X-ray irradiation. Technology computer-aided design (TCAD) simulation method was also employed to explore the underlying physical mechanisms. The results demonstrate that the combined effect of \hl{a positive pulse injected into the collector (CEMP)} and pulsed X-ray irradiation exceeds the linear superposition of their individual effects. Conversely, the combined effect of \hl{a positive pulse injected into the base (BEMP)} and pulsed X-ray irradiation aligns closely with the results observed under BEMP acting alone. Mechanism analysis reveals that when CEMP and pulsed X-ray irradiation act simultaneously, there is a significant increase in both the drift photocurrent at the collector junction and the diffusion photocurrent near the collector junction. However, when BEMP and pulsed X-ray irradiation act simultaneously, these photocurrent components remain small, leading to a combined effect similar to the results observed when BEMP acts alone. These findings provide critical insights for the radiation-hardening design of bipolar circuits in harsh radiation environments.
\end{abstract}


\section{Introduction}
\label{sec:introduction}
\begin{figure}
  \centerline{\includegraphics[width=2.8in]{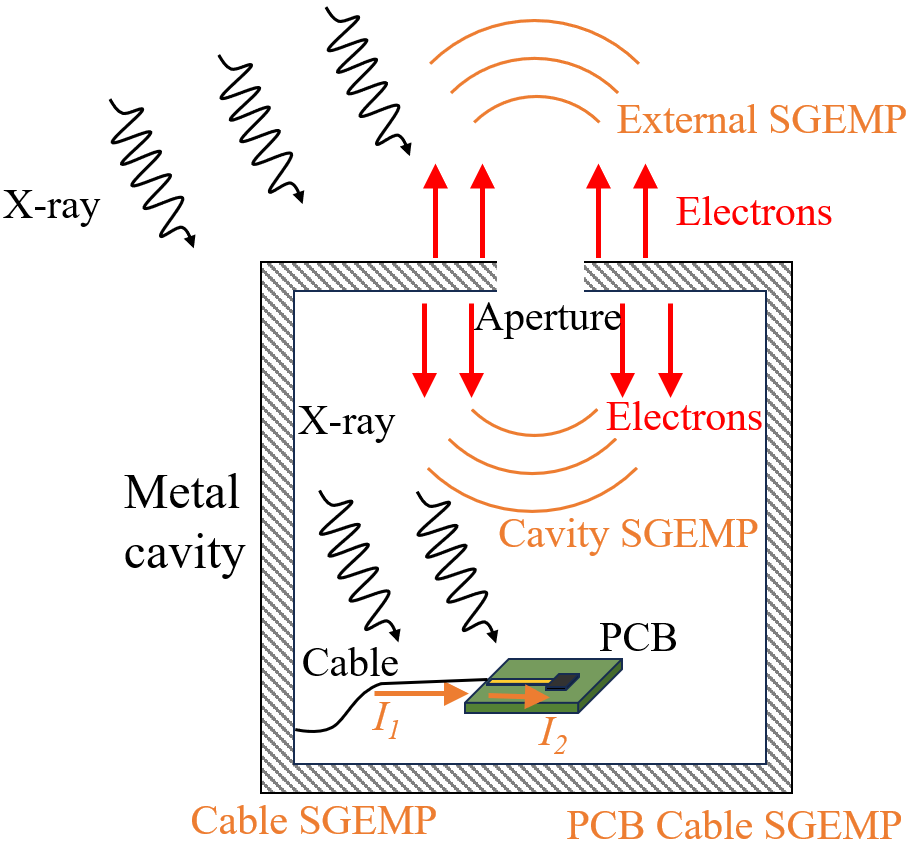}}
  \caption{Schematic diagram of various types of SGEMP generated during transient X-ray irradiation of electronic equipment.}\label{intro_zmx}
\end{figure}

\IEEEPARstart{I}{ntense} nanosecond-level transient X-rays are present in physical experiments and applications involving transient X-ray radiation, such as within the target chambers of high-power laser inertial confinement fusion (ICF) facilities \cite{fournier2015nif}. When such intense transient X-rays irradiate electronic equipment, high-speed electrons are excited within attoseconds through the photoelectric effect and compton scattering \cite{ossiander2018absolute}. These excited high-speed electrons immediately begin moving, forming a transient space current that directly stimulates the system-generated electromagnetic pulse (SGEMP), the initiation of which is nearly synchronous with the moment X-rays irradiate the material and generate electrons \cite{xu2022simulation, xu2017evaluation, xu2021code, zhang2022anti, zhang2024calculation, xu2022thesis}. As shown in Fig.~\ref{intro_zmx}, the SGEMP generated by transient X-ray irradiation of electronic equipment includes: external SGEMP stimulated by electrons emitted outward from metal cavities \cite{xu2022simulation}; cavity SGEMP stimulated by electrons emitted inward from metal cavity walls \cite{xu2021code}; cable SGEMP stimulated by electrons emitted from the metal shield and core wire of cables \cite{xu2017evaluation,zhang2022anti}; and PCB cable SGEMP stimulated by electrons emitted from microstrip lines on printed circuit boards (PCBs) \cite{zhang2024calculation}. The rise time of these SGEMPs typically closely follows the rise time of the X-ray pulse, but the fall time and total duration often exceed those of the X-ray pulse itself, resulting in significant temporal overlap, particularly during the rising edge and peak period \cite{zhang2024calculation, xu2022thesis}.

Both X-rays and electromagnetic pulses propagate at the speed of light. When X-rays irradiate semiconductor devices, large number of excess carriers within the silicon material are generated. The drift and diffusion movements of these carriers form a photocurrent, inducing transient ionizing radiation effects (TREEs) and causing current and voltage perturbations in the device \cite{li2022investigation, li2022experimental}. Simultaneously, various SGEMPs can inject transient pulses into the device pins through radiative and conductive coupling paths, also inducing current and voltage disturbances within the device \cite{stathis1994radiation, consoli2020laser, brown2013analysis, brown2012assessment, Batani2023Europe}. Electric fields as high as MV/m levels have been observed inside laser ICF facilities, posing a serious threat to electronic equipment placed inside and outside the laboratory \cite{falcone2019grant}. EMP research is considered a crucial subject and is one of the three key research areas in high-intensity laser system experimental technology, as the anticipated EMP environment in future laser facilities will far exceed any currently experienced globally \cite{falcone2019grant}. Therefore, it is essential to study the combined effects when X-rays and EMP act simultaneously on devices. The fundamental circuit units of modern electronic systems primarily consist of bipolar junction transistors (BJTs) and metal-oxide-semiconductor field-effect transistors (MOSFETs). Studying the combined effects on individual BJTs is crucial for understanding complex circuit behaviors.

In previous work, the mechanisms of the independent TREE \cite{carr1964transient, wirth1964transient, fjeldly2001modeling, alexander2003transient} and EMP \cite{camp2002influence, garrett2013impact} on BJTs have been thoroughly investigated. In recent years, researchers mainly focused on the combined effects of total ionizing dose and transient ionizing \cite{li2019synergistic,ruibin2018impact, yan2020accumulative}, the combined effects of displacement and transient ionizing \cite{wang2022photocurrent, wang2023displacement, roig2014impact}, and the combined effects of the total ionizing dose and electromagnetic pulse \cite{doridant2012impact, estep2012electromagnetic, wu2022synergistic, lawal201860co, wu2023combined}. However, published work on the combined effects of transient ionization and EMP remains relatively limited.

D. H. Habing, et al\cite{habing1970response, hartman1975electrical} investigated the effects of transient ionizing radiation on the pulsed high current failure levels of semiconductor junctions, with injected pulse power up to several hundred watts. However, they only conducted experimental studies without simulation analysis. M. Zhao et al.\cite{mo2017combined} performed numerical simulations on the comprehensive impact of EMP and transient ionizing radiation on bipolar transistors. Their simulation results indicated that the simultaneous action of EMP and transient ionizing radiation is more likely to damage bipolar transistors than their individual actions. Nevertheless, that study only involved simulations and lacked experimental validation. Furthermore, these studies primarily focused on the burnout effects of simultaneous transient X-ray and EMP action on BJTs, while the transient response under combined effects remains largely unexplored. Our team previously conducted preliminary simulation studies on the transient response of bipolar transistors subjected to simultaneous transient ionizing radiation and EMP \cite{zhong2024two}. 

This paper experimentally investigates the combined effects of transient ionizing radiation and EMP on vertical NPN bipolar transistors and employs technology computer-aided design (TCAD) simulations to deeply analyze the physical mechanisms behind these combined effects. In practical scenarios, EMP can occur simultaneously on multiple electrodes of a transistor, such as the collector, base, and emitter. However, the amplitude of the EMP on each electrode depends on numerous factors: the position of the electronic system within the cavity, the characteristics (type and length) of connecting cables, and the interconnection topology between the transistor and other components or even the entire circuit. Given this complexity, this study adopts a controlled-variable approach, starting with the most basic synergistic effects of X-ray and EMP injection into a single electrode. This aims to clarify the effects of individual variables and establish an important experimental and theoretical foundation for future multi-electrode combined effects research. The findings can provide valuable insights for the radiation-hardening design of bipolar circuits in harsh radiation environments.

\begin{figure*}
  \centerline{\includegraphics[width=6.2in]{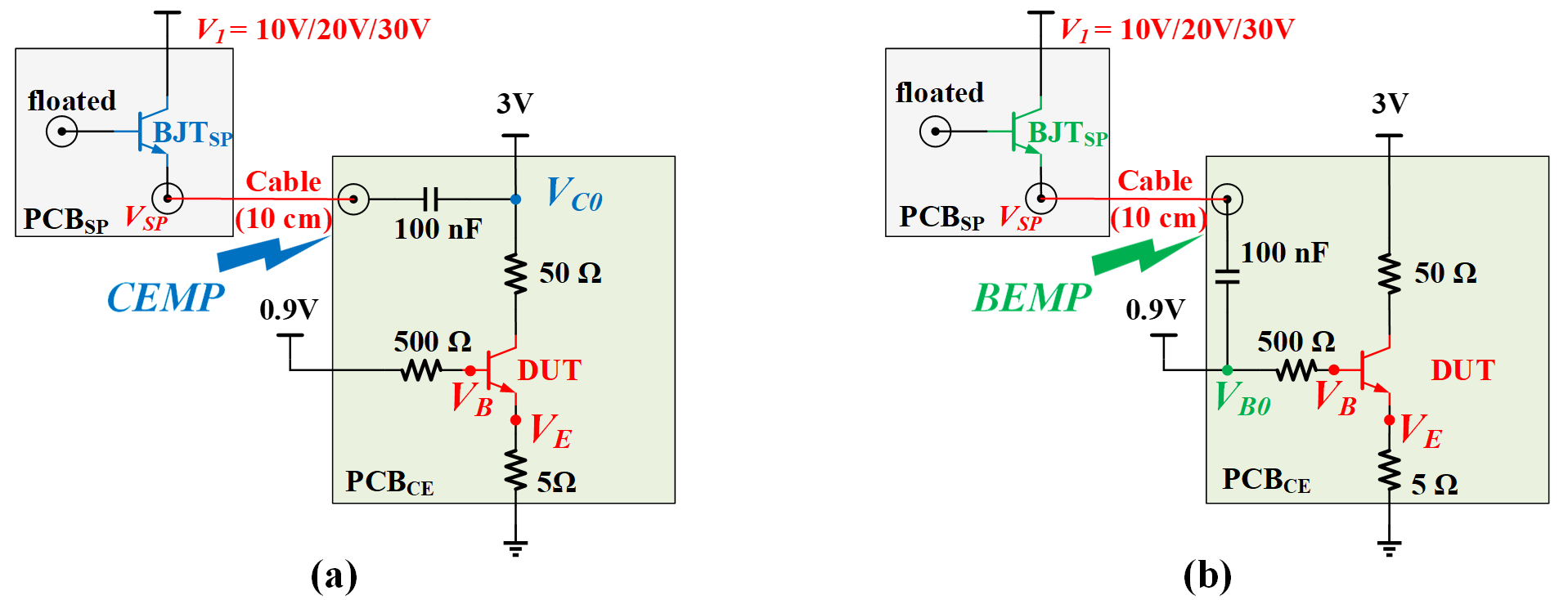}}
  \caption{Bias configurations of the 2N2222A transistor under pulsed X-ray irradiation: (a) Combined effect of pulsed X-ray irradiation and \hl{pulse injected into the collector (CEMP)}; (b) Combined effect of pulsed X-ray irradiation and \hl{pulse injected into the base (BEMP)}.}\label{testbench}
\end{figure*}

\section{Experimental Details and Results}

\subsection {Experimental Details}

The 2N2222A transistor manufactured by Microsemi Corporation (USA) was selected as the device under test(DUT), due to its widespread use and rich literature support in radiation effects research \cite{li2019synergistic, wu2023combined, binghuang2020tcad, nationsa2008, verley2012}. To minimize sample variability, all tested samples were selected from the same production batch (identical date code) and exhibited consistent electrical parameters (current gain). 

To experimentally investigate the combined effects of transient ionization and EMP, we conducted pulsed X-ray irradiation experiments on transistors at the "Shenguang II" accelerator \cite{dingguo2021intense}. The pulsed X-rays generated by this accelerator cover an area of $500\,\text{cm}^{\text{2}}$, with a typical full width at half maximum (FWHM) of 50\,ns, an average energy of 87\,keV, an average energy fluence of $36\,\text{mJ/cm}^{\text{2}}$${V}_{{B}}$, and a dose uniformity ratio of 2:1.

Fig.~\ref{testbench} shows the bias configuration of transistors under pulsed X-ray irradiation. Two types of test boards were used in the experiment: the secondary photocurrent test board ($\text{PCB}_{\text{SP}}$) and the combined effect test board ($\text{PCB}_{\text{CE}}$). We innovatively utilized the photocurrent response of the 2N2222A transistor ($\text{BJT}_{\text{SP}}$) on the $\text{PCB}_{\text{SP}}$ board under X-ray irradiation as an EMP source.
\hl{A pulse was injected into the collector} (see Fig.~\ref{testbench}(a)) or base (see Fig.~\ref{testbench}(b)) of the DUT(i.e., the BJT on $\text{PCB}_{\text{CE}}$) via a 0.1\,m coaxial cable and a 100\,nF capacitor. Given the 0.1\,m  coaxial cable between $\text{BJT}_{\text{SP}}$ and the DUT, the time delay between the voltage pulse and the pulsed X-ray irradiation was only about 0.3\,ns, less than 0.7\% of the 50\,ns FWHM of the pulsed X-ray irradiation. Therefore, in the experiment, the pulsed X-ray irradiation and the voltage pulse can be considered to act on the DUT simultaneously.This innovative design effectively circumvented the technical challenge of precisely synchronizing transient X-ray and EMP triggering. Furthermore, this design achieved an amplitude-adjustable EMP source within a very small space (the volume of the 2N2222A and $\text{PCB}_{\text{SP}}$ is very small), allowing the placement of multiple transistors in a single shot experiment to simultaneously measure the individual and synergistic effects of transient X-ray and EMP.

As shown in Fig.~\ref{testbench}, the base of $\text{BJT}_{\text{SP}}$ is floated, the collector is connected to a DC power supply (${V}_{{1}}$), and the emitter is connected to the collector or base of the DUT. Under pulsed X-ray irradiation, secondary photocurrent is generated within $\text{BJT}_{\text{SP}}$, causing transient voltage variations at the emitter load of $\text{BJT}_{\text{SP}}$ (${V}_{{SP}}$). This phenomenon, known as the secondary photocurrent response of bipolar transistors under pulsed X-ray irradiation \cite{carr1964transient, wirth1964transient, fjeldly2001modeling, alexander2003transient} , was utilized to apply voltage pulses to the collector or base of the DUT via a coaxial cable and a capacitor. By adjusting the amplitude of ${V}_{{1}}$, we can change the amplitude of ${V}_{{SP}}$, consequently, the amplitude of \hl{the pulses injected into the DUT.}

The DUT was configured in a forward-active bias arrangement: the collector was connected to a 3\,V DC supply via a 50\,$\Omega$ resistor, the base was connected to a 0.9\,V DC supply via a 500\,$\Omega$ resistor, and the emitter was grounded via a 5\,$\Omega$ resistor. As shown in Fig.~\ref{testbench}(a), when studying the combined effect of pulsed X-ray irradiation and \hl{pulses injected into the collector (CEMP)}, we measured the voltage waveform between the 100\,nF coupling capacitor and the 50\,$\Omega$ collector load ($V_{C0}$), the transient voltage response at the base (${V}_{{B}}$), and the transient voltage response at the emitter ($V_{E}$) using oscilloscopes in high-impedance mode. As shown in Fig.~\ref{testbench}(b), when studying the combined effect of pulsed X-ray irradiation and \hl{pulses injected into the base (BEMP)}, we measured the voltage waveform (${V}_{{B0}}$) between the 100\,nF coupling capacitor and the 500\,$\Omega$ base load, the transient voltage response at the base (${V}_{{B}}$), and the transient voltage response at the emitter (${V}_{{E}}$).

We conducted one preliminary experiment and six combined-effects experiments. In each experiment, all samples were placed at equal distances from the accelerator target to ensure a uniform peak dose rate of approximately \num{1.0e9} rad(Si)/s. The preliminary experiment aimed to measure the secondary photocurrent response of the $\text{BJT}_{\text{SP}}$ under different bias voltages (${V}_{{1}}$), which would be used to \hl{inject pulses into the DUT} in subsequent combined effects experiments. In this experiment, we used three pristine transistors and three $\text{PCB}_{\text{SP}}$ boards. The bases of the three transistors were all floated, and ${V}_{{1}}$ was set to 30\,V, 20\,V, and 10\,V, respectively (voltage waveforms were measured using a high-impedance oscilloscope). The emitters were directly connected to a 50\,$\Omega$ termination load oscilloscope for measuring the transient response of ${V}_{{SP}}$ \cite{li2019synergistic,wang2022photocurrent}. 

\begin{table}[!t]
\caption{specific settings for the combined-effects experiments}
\label{table_setting}
\centering
\begin{tabular}{ c c c c }
\hline
Combined-Effects Experiment  & DUT1             & DUT2          & DUT3   \\
\hline
Experiments 1-3        & X-ray + CEMP      & X-ray        &  CEMP   \\

Experiments 4-6        & X-ray + BEMP      & X-ray        &  BEMP   \\
\hline
\end{tabular}
\end{table}

Table \ref{table_setting} lists the specific settings for the combined-effects experiments. Experiments 1-3 used the configuration shown in Fig.~\ref{testbench}(a), while experiments 4-6 used the configuration shown in Fig.~\ref{testbench}(b). In each combined-effects experiment, we used five pristine transistors, two $\text{PCB}_{\text{SP}}$ settings, and three $\text{PCB}_{\text{CE}}$ settings. Three transistors served as DUTs, while the remaining two were used to \hl{inject pulses into the DUTs}. Among the three DUTs, one was exposed to both pulsed X-ray irradiation and the voltage pulse, one was shielded with 2 \,mm thick lead and thus only exposed to the voltage pulse, and the third was not connected to the $\text{PCB}_{\text{SP}}$ and thus only exposed to pulsed X-ray irradiation. A total of 30 transistors were used in the combined-effects experiments.

\subsection {Experimental Results}

\begin{figure}
  \centerline{\includegraphics[width=3.1in]{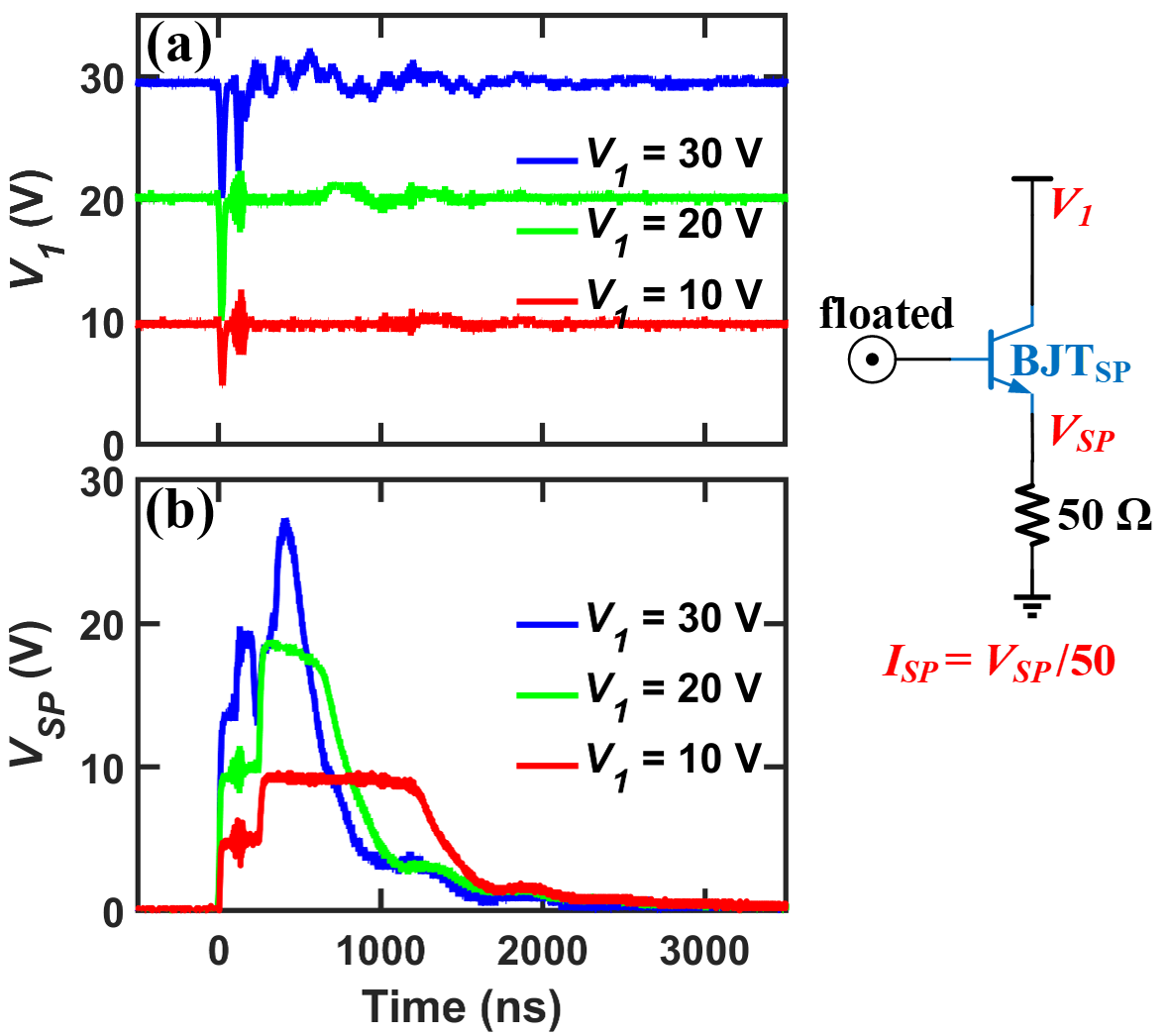}}
  \caption{Experimental results of the secondary photocurrent response (${V}_{{SP}}$) under different biases (${V}_{{1}}$): (a) Waveform of ${V}_{{1}}$; (b) Transient voltage response at the emitter (${V}_{{SP}}$).}\label{mea_input}
\end{figure}

 Fig.~\ref{mea_input} presents the experimental results of the secondary photocurrent response of the $\text{BJT}_{\text{SP}}$ under different bias voltages (${V}_{{1}}$), which were used to \hl{inject pulses into the DUT} in the subsequent combined-effects experiments. The first observed characteristic is that the duration of the secondary photocurrent (${V}_{{SP}}$) response exceeds 1\,µs, significantly greater than the 50\,ns FWHM of the pulsed X-ray, consistent with the results reported in the literature\cite{li2019synergistic}. This phenomenon can be attributed to the base-floated bias condition, where the base is floated and the collector junction is reverse-biased. Under pulsed X-ray irradiation, carriers generated within the collector junction drift under the electric field. Electrons in the base region diffuse toward the junction, and holes within a diffusion length in the collector region reach the collector junction and enter the base, causing a significant increase in hole concentration in the base. This lowers the barrier between the base and emitter, causing the emitter junction to become forward-biased. A large number of electrons from the emitter enter the base, with a minority recombining with holes, while the majority diffuse to the collector junction and are collected by the electric field, forming the diffusion current. When the minority carrier lifetime within the device exceeds the duration of the pulsed X-ray, the duration of the secondary photocurrent will exceed that of the pulsed X-ray, consistent with the conclusions drawn from previous studies \cite{carr1964transient, wirth1964transient, fjeldly2001modeling, alexander2003transient}.

The second observed characteristic is that the higher the collector bias voltage ${V}_{{1}}$, the larger the peak of ${V}_{{SP}}$. This is because an increase in ${V}_{{1}}$ leads to an increase in the depletion region width, significantly enhancing the drift current component of the secondary photocurrent. The third characteristic is that the lower the collector bias voltage ${V}_{{1}}$, the longer the saturation plateau duration of the ${V}_{{SP}}$ waveform, but the total duration remains essentially unchanged. This is because a decrease in ${V}_{{1}}$ reduces the amplitude of the saturation photocurrent (${I}_{{SP}}$ = ${V}_{{SP}}/50\,\Omega$, limited by the low bias voltage), requiring more time to collect an equivalent amount of charge. Consequently, as ${V}_{{1}}$ decreases, the plateau period of the diffusion current component of the secondary photocurrent lengthens. However, the total duration of the secondary photocurrent is primarily determined by the minority carrier lifetime ( $\approx$\,3000 ns) and is not affected by the collector bias voltage ${V}_{{1}}$.

\begin{figure*}
  \centerline{\includegraphics[width=7.0in]{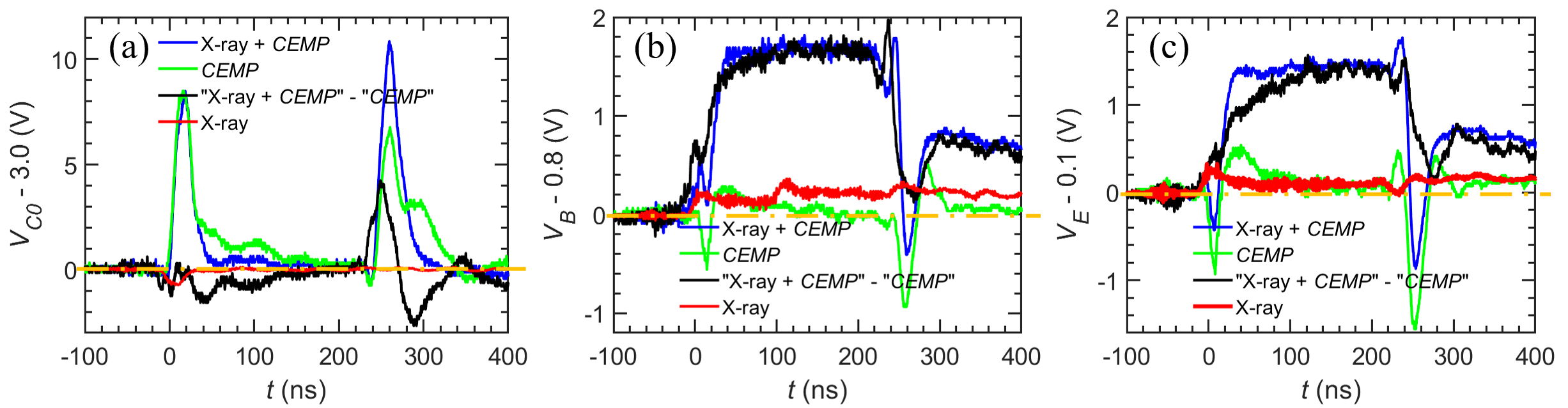}}
  \caption{Experimental results of the combined effects of pulsed X-ray irradiation and CEMP ($V_{1}$ = 30\,V): (a) Waveform of ${V}_{{C0}}$; (b) Base voltage response (${V}_{{B}}$); (c) Emitter voltage response (${V}_{{E}}$).}\label{mea_coll}
\end{figure*}

\begin{figure*}
  \centerline{\includegraphics[width=7.0in]{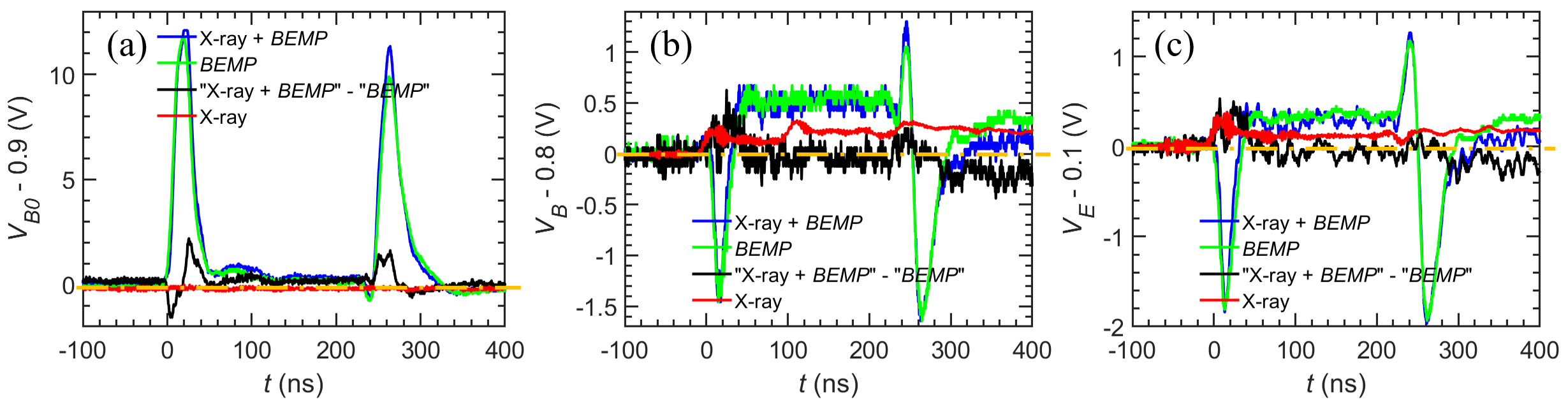}}
  \caption{Experimental results of the combined effects of pulsed X-ray irradiation and BEMP ($V_{1}$ = 30\,V): (a) Waveform of ${V}_{{B0}}$; (b) Base voltage response (${V}_{{B}}$); (c) Emitter voltage response (${V}_{{E}}$).}\label{mea_base}
\end{figure*}

Furthermore, due to the absence of a large bypass capacitor at ${V}_{{1}}$ power terminal, the ${V}_{{SP}}$ waveform exhibits a two-step pattern with an interval of approximately 250\,ns between steps, and the entire photocurrent lasts for over 1\,µs. These two steps \hl{inject two pulses into the DUT}, each with a half-width of about 30\,ns and separated by about 250\,ns, as shown in Fig.~\ref{mea_coll}(a) and Fig.~\ref{mea_base}(a). Therefore, the first pulse acts on the DUT simultaneously with the pulsed X-ray irradiation, while the second pulse acts on the DUT after the pulsed X-ray irradiation has ended. Compared to \hl{injecting only a single pulse}, this experimental design allows us to observe not only the effect of the simultaneous action of the voltage pulse and X-ray pulse but also the response \hl{when an additional pulse is injected later}. Moreover, the subsequently injected second pulse does not affect the measurement of the response peak of the first pulse. In practical applications, whether there are two voltage pulses depends on the interconnections with other transistors or chips of the entire electronic system, as well as the position of the electronic system in the cavity and the length of the cables connected to it.

\begin{figure}
  \centerline{\includegraphics[width=2.8in]{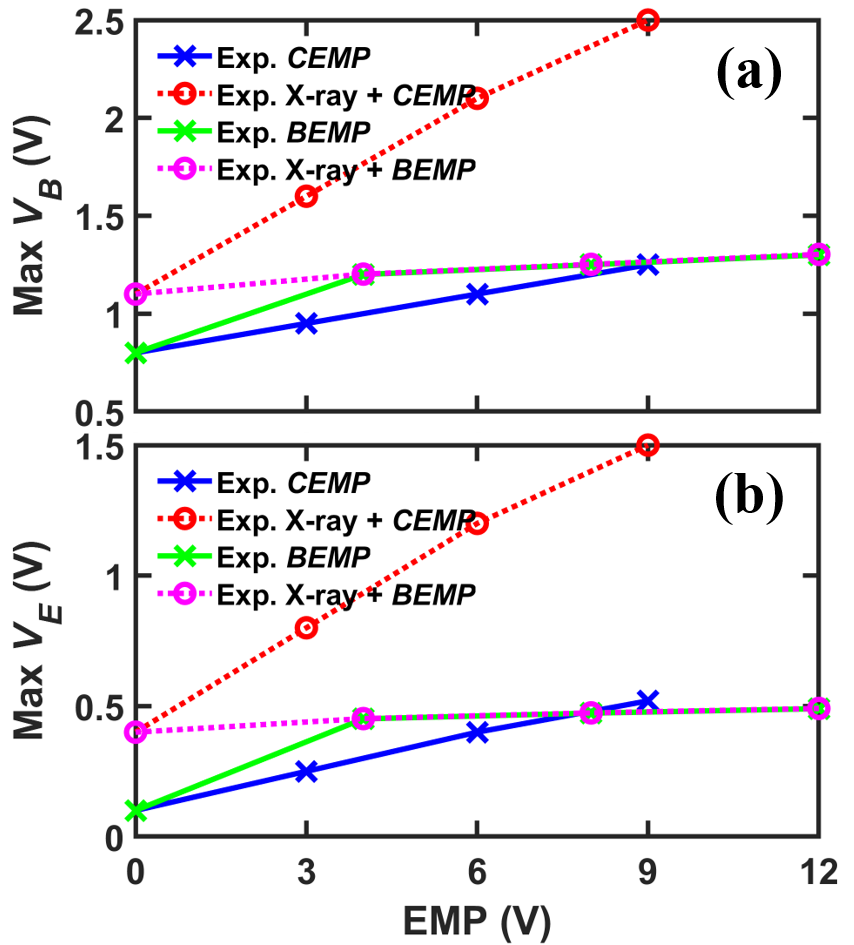}}
  \caption{Experimental results of the combined effects of pulsed X-ray irradiation and voltage pulse with varying amplitudes: (a) Maximum base voltage response (0-200\,ns); (b) Maximum emitter voltage response (0-200\,ns).}\label{mea_base_diff}
\end{figure}

Fig.~\ref{mea_coll} shows the experimental results of the combined effect of pulsed X-ray irradiation and CEMP, with the $V_{1}$ of $\text{BJT}_{\text{SP}}$ set at 30\,V. Fig.~\ref{mea_coll}(a) shows the ${V}_{{C0}}$ waveform of the three DUTs, reflecting \hl{the pulse waveforms injected into the DUTs}. It can be observed that the blue waveform (pulsed X-ray irradiation and CEMP acting simultaneously) and the green waveform (only CEMP acting) almost coincide, indicating that \hl{the pulses injected into the DUT collector }under these two conditions are similar. Meanwhile, the red waveform (only pulsed X-ray irradiation acting) exhibits only a small negative perturbation, indicating that this DUT remains in its steady-state DC bias without any \hl{injected pulse} and was only affected by X-ray irradiation. Fig.~\ref{mea_coll}(b) and~\ref{mea_coll}(c) show the base and emitter voltage responses of the three DUTs, respectively. The measured combined effect is the key finding of this study.

As shown in Fig.~\ref{mea_coll}(b) and~\ref{mea_coll}(c), when the transistor operates normally, the base voltage is approximately 0.8\,V and the emitter voltage is approximately 0.1\,V. When exposed to pulsed X-ray irradiation alone, both the base and emitter exhibit positive voltage perturbations lasting over 400\,ns, with the maximum perturbation of approximately 0.3\,V at both the base and emitter. When only CEMP is injected, the voltage responses at the base and emitter initially show a brief negative perturbation, followed by a sustained positive perturbation, with a maximum positive perturbation of about 0.45\,V at the base and 0.42\,V at the emitter. Simultaneous exposure to pulsed X-ray irradiation and CEMP causes a brief negative perturbation at the base and emitter, which then transitions to a significant positive perturbation, reaching about 1.7\,V at the base and 1.4\,V at the emitter. The black waveform in Fig.~\ref{mea_coll}(b) and~\ref{mea_coll}(c) is the result of subtracting the green waveform from the blue waveform (i.e., combined effect minus CEMP-only effect). It can be seen that the black waveform is much larger than the red waveform (X-ray-only effect), indicating that the combined effect of pulsed X-ray irradiation and CEMP exceeds the linear superposition of their individual effects.

Fig.~\ref{mea_base} shows the experimental results of the combined effects of pulsed X-ray irradiation and BEMP, with the ${V}_{{1}}$ of $\text{BJT}_{\text{SP}}$ set at 30\,V. Fig.~\ref{mea_base}(a) shows the ${V}_{{B0}}$ waveform of the three DUTs. As shown in Fig.~\ref{mea_base}(b) and~\ref{mea_base}(c), when pulsed X-ray irradiation acted alone, both the base and emitter exhibited positive voltage perturbations lasting over 400\,ns, with the maximum perturbations of approximately 0.3\,V both at the base and emitter. When only BEMP was applied, the voltage responses at the base and emitter initially showed a brief negative perturbations, followed by sustained positive perturbations. The maximum positive perturbations reached approximately  1.3\,V at the base and 0.49\,V at the emitter. Simultaneous exposure to pulsed X-rays and BEMP resulted in initial brief negative perturbations at both the base and emitter, which then transitioned to sustained positive perturbations with comparable amplitudes to those observed when BEMP acted alone. The black waveform (combined effect minus BEMP-only effect) in Fig.~\ref{mea_base}(b) and~\ref{mea_base}(c) is close to zero, indicating that the combined effect of pulsed X-ray irradiation and BEMP closely resembles the effect of BEMP acting alone.

Subsequent analysis revealed that the negative pulse segments in the time ranges of 0–50\,ns and 250–300\,ns in the combined effect (blue line) and EMP response (green line) for the base in Figs.~\ref{mea_coll}(b) and Figs.~\ref{mea_base}(b) and emitter in Figs.~\ref{mea_coll}(c) and Figs.~\ref{mea_base}(c) are due to crosstalk during the injection pulse, superimposing a negative pulse on the pulse response of 2N2222A. This is because the experimental system is relatively complex, involving factors such as coupling capacitance and 30-meter-long test cables, and the signal amplitudes at the base and emitter are about an order of magnitude smaller than the injected pulse signal.

Therefore, Fig.~\ref{mea_base_diff} summarizes the maximum positive perturbations within the 0-200\,ns time range for (a) base voltage and (b) emitter voltage under different voltage pulse amplitudes, reflecting the experimental trends of the combined effects of pulsed X-ray irradiation and voltage pulses of varying amplitudes.The results show that the trends of the combined effects observed under peak voltage pulses of 3\,V, 6\,V, and 9\,V are consistent: the combined effect of pulsed X-ray and CEMP exceeds the linear superposition of their individual effects, while the combined effect of pulsed X-ray and BEMP is very similar to the effect of BEMP alone. It is noteworthy that as the peak CEMP amplitude increases, the degree to which its combined effect with pulsed X-ray exceeds linear superposition becomes more significant.

\section{TCAD SIMULATION AND MECHANISM ANALYSIS }
\subsection{Simulation Method}

\begin{figure}
  \centerline{\includegraphics[width=3.5in]{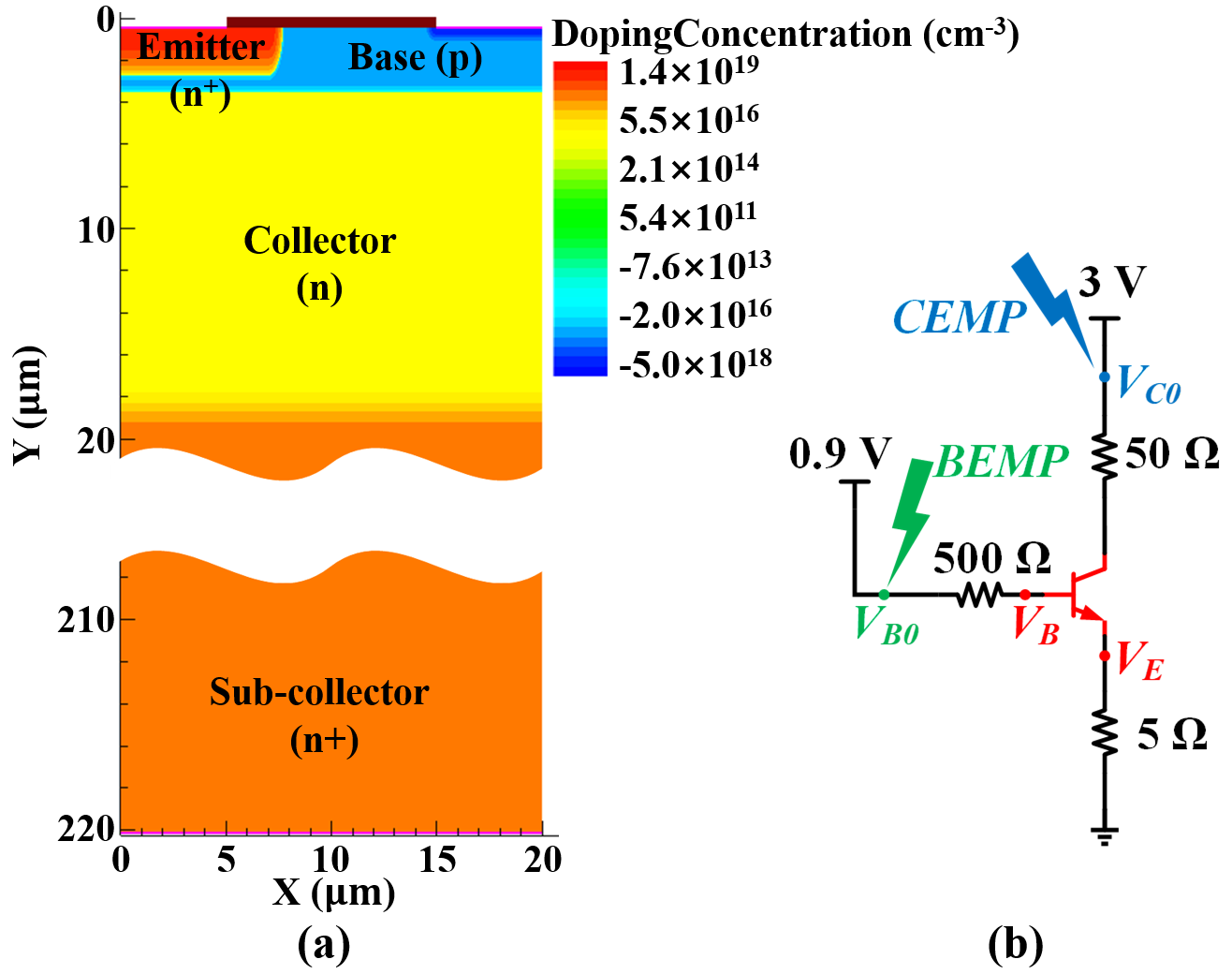}}
  \caption{(a) 2-D model of the 2N2222A transistor; (b) Schematic diagram of circuit connections for TCAD simulation.}\label{tcad_model}
\end{figure}

Fig.~\ref{tcad_model} presents the 2-D model of the 2N2222A transistor and the schematic diagram of circuit connections for TCAD simulation. As shown in Fig.~\ref{tcad_model}(a), the 2N2222A is a substrate transistor, epitaxially grown on a heavily doped N-type substrate \cite{li2019synergistic, wu2023combined, binghuang2020tcad, nationsa2008, verley2012}.  The upper surface features the emitter on the left and the base on the right, with the substrate serving as the collector. The epitaxial layer of collector and the base form a graded p-n junction capable of withstanding high reverse bias. Below the epitaxial layer lies a heavily doped region extending to the bottom of the collector, with the collector situated on the backside of this region. Table \ref{table_edoping} lists the specific dimensions and doping concentrations of each region of the transistor \cite{li2019synergistic, wu2023combined, binghuang2020tcad, nationsa2008, verley2012}. The peak doping concentration of the emitter is approximately \num{1.4e19}\,\si{cm^{-3}}. The peak doping concentrations of the base and the base epitaxial layer are \num{1.4e18}\,\si{cm^{-3}} and \num{1.0e17}\,\si{cm^{-3}}, respectively. The doping concentrations in the heavily doped region and the epitaxial region of the collector are \num{1.0e18}\,\si{cm^{-3}} and \num{1.0e16}\,\si{cm^{-3}}, respectively. The thickness of the emitter, base, collector, and sub-collector are 2.0, 0.6, 17.0, and 200.0\,µm, respectively. 

\begin{table}[!t]
\caption{Specific dimensions and doping concentrations of the 2N2222A transistor}
\label{table_edoping}
\centering
\begin{tabular}{ c c c }
\hline
 Region         & Doping (\si{cm^{-3}}) & Depth (\si{\micro\meter}) \\
\hline
Emitter($n^+$) & \num{1.4e19}        & 2.0 \\

Base(p)        & \num{1.0e17}       & 0.6 \\

Collector(n)   & \num{1.0e16}        & 17  \\

Sub-collector($n^+$) & \num{1.0e18}  & 200 \\
\hline
\end{tabular} 
\end{table}

Compared with the 3D simulations, 2D-TCAD simulations exhibit better convergence and faster computational speeds. Many scholars have employed 2D-TCAD simulations for mechanism analysis \cite{li2019synergistic, wu2023combined, binghuang2020tcad}.
This study employs 2D-TCAD simulation to analyze the combined effects, with more detailed 3D modeling planned for future work. Our previous simulation study \cite{zhong2024two} did not account for the series load resistors used in this experiment, and the voltage pulse half-widths and amplitudes used also differed significantly from those in this experiment. Therefore, to enable more precise mechanism analysis, new 2D-TCAD simulations were performed for this paper.

\begin{figure}
  \centerline{\includegraphics[width=3.0in]{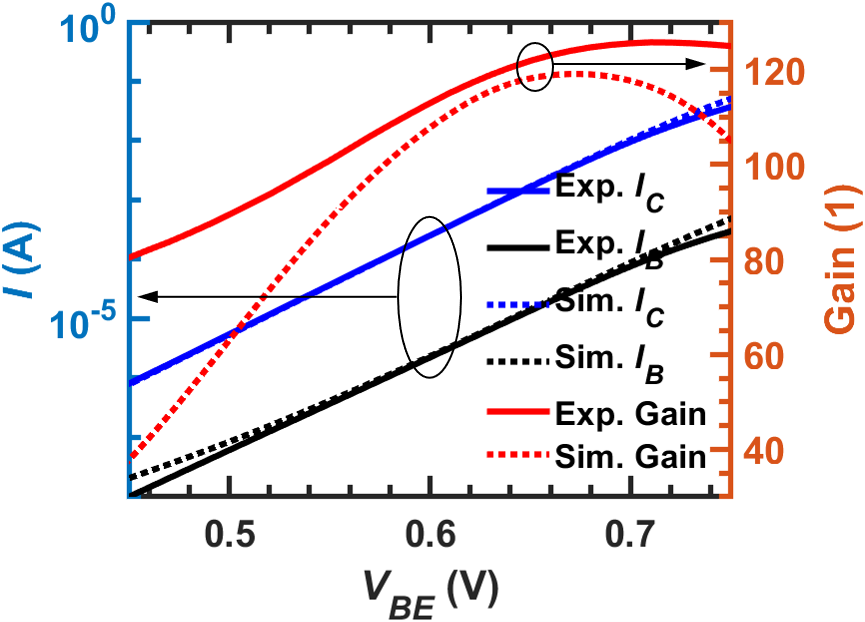}}
  \caption{Experimental and TCAD simulation results of Gummel characteristic curve. Note: Current (left axis) uses a logarithmic scale, while Gain (right axis) uses a linear scale.}\label{Gummel}
\end{figure}

Fig.~\ref{Gummel} presents the measured and simulated results of Gummel characteristic curve (${I}_{{C}}$, $I_{B}$ and Gain $\beta$= $I_{C}$/$I_{B}$ versus $V_{BE}$). The emitter was grounded, the collector–emitter voltage $V_{CE}$ was fixed at 1\,V, and the base–emitter voltage $V_{BE}$ was swept from 0.4 to 0.8\,V. As shown in Fig.~\ref{Gummel}, the simulation results of $I_{C}$ are in good agreement with the experimental results. While there are some differences between the simulation and experimental results of $I_{B}$ at low and high $V_{BE}$ values, the error between the simulation and experimental results for the gain remains within 50\,\%  (minimum error of 2.5\% at $V_{BE}$ = 0.63\,V, simulated $\beta$ = 115, experimental $\beta$ = 118; maximum error of 50\% at $V_{BE}$=0.4 V, simulated $\beta$ = 40, experimental $\beta$ = 80).

\begin{figure}
  \centerline{\includegraphics[width=2.6in]{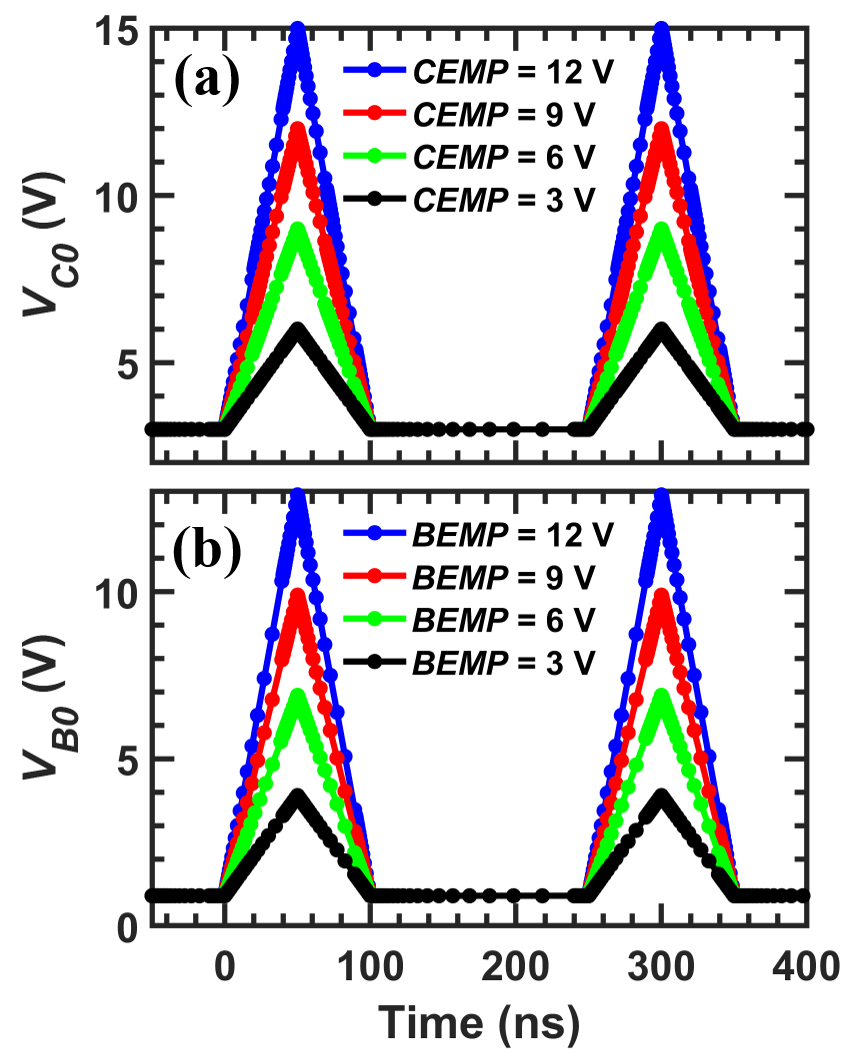}}
  \caption{Idealized voltage pulse waveforms used in the simulation: (a) $V_{C0}$ waveform when \hl{applied to} the collector; (b) $V_{B0}$ waveform when \hl{applied to} the base.}\label{tcad_pulse}
\end{figure}

As shown in Fig.~\ref{tcad_model}(b), the circuit connections for TCAD simulation involved biasing the transistor in the forward amplification mode ($V_{E0}$ = 0\,V, $V_{C0}$ = 3\,V, $V_{B0}$ = 0.9\,V), with a series resistance of 5\,$\Omega$ at the emitter, 500\,$\Omega$ at the base, and 50\,$\Omega$ at the collector. Due to the complexity of the \hl{pulse waveforms in the experiments}, we simplified them to ideal triangular pulses in the simulations, as shown in Fig.~\ref{tcad_pulse}. The point markers in Fig.~\ref{tcad_pulse} reflect the time step in the simulation, with a minimum step size of 0.1\,ns. Two triangular voltage pulses with a rise time of 50\,ns and a fall time of 50\,ns are \hl{applied as a boundary condition to} either the base or the collector, with a 250\,ns interval between the two pulses, and the transient voltage and current responses of each electrode are computed. The peak values of the voltage pulses are set at 3\,V, 6\,V, 9\,V and 12\,V. Specifically, Fig.~\ref{tcad_pulse}(a) illustrates the waveform of $V_{C0}$ when the voltage pulses are \hl{applied to} the collector, while Fig.~\ref{tcad_pulse}(b) illustrates the waveform of $V_{B0}$ when the voltage pulses are \hl{applied to} the base.

The underlying photon-response mechanism in semiconductors involves the generation of excess carriers due to energy absorption in the silicon material. The generation coefficient for excess carriers in silicon is \num{4e13} h–e pairs/\si{cm^3}rad(Si) \cite{alexander2003transient}. Consequently, the corresponding number of excess carriers is introduced into the bulk silicon of the transistor to simulate the photocurrent \cite{li2019synergistic}. In the simulation, the transient ionizing radiation of the transistor is set to a peak dose rate of \num{1e9} rad(Si)/s, with the waveform defined as a Gaussian function with a standard deviation of 50\,ns, consistent with the experimental conditions.

\subsection{Simulation Results}

\begin{figure*}
  \centerline{\includegraphics[width=7.0in]{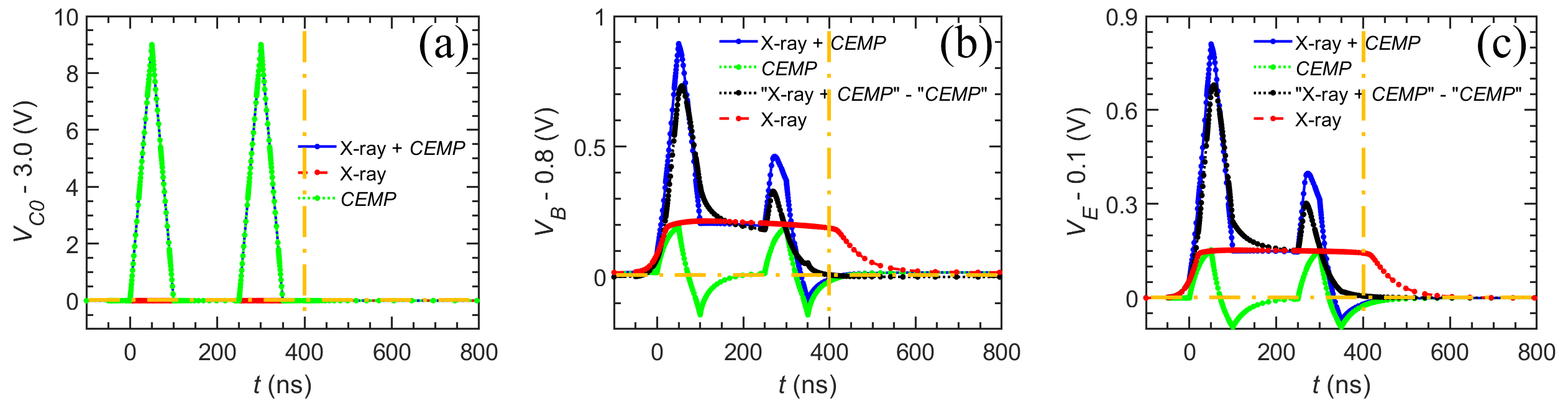}}
  \caption{Simulation results of the combined effects of pulsed X-ray irradiation and CEMP with a peak value of 9\,V: (a) Waveform of $V_{C0}$; (b) Base voltage response (${V}_{{B}}$); (c) Emitter voltage response (${V}_{{E}}$).}\label{sim_coll}
\end{figure*}

\begin{figure*}
  \centerline{\includegraphics[width=7.0in]{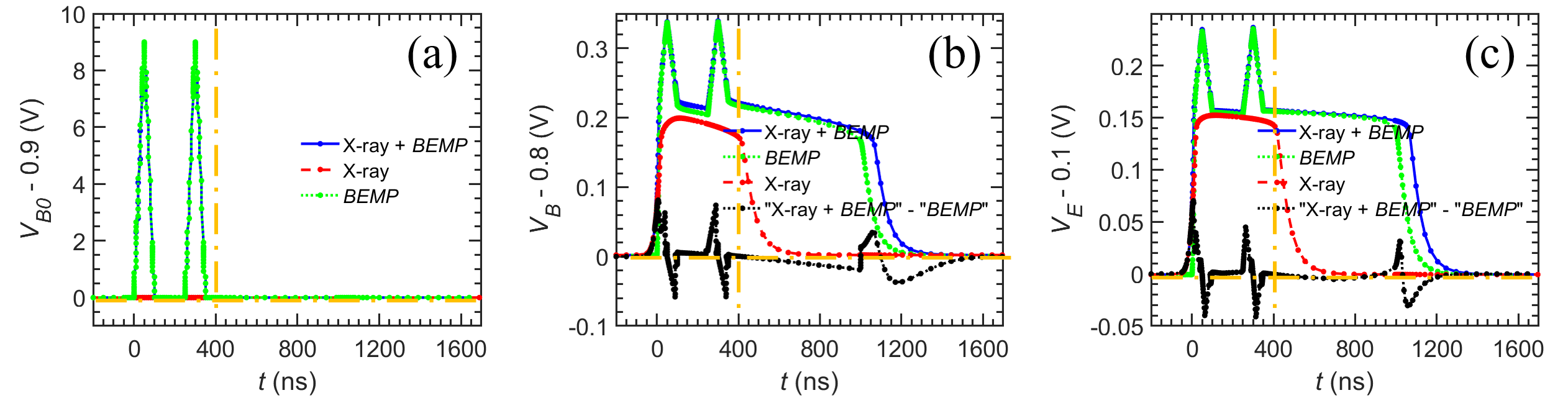}}
  \caption{Simulation results of the combined effects of pulsed X-ray irradiation and BEMP with a peak value of 9\,V: (a) Waveform of $V_{B0}$; (b) Base voltage response (${V}_{{B}}$); (c) Emitter voltage response (${V}_{{E}}$).}\label{sim_base}
\end{figure*}

\begin{figure}
  \centerline{\includegraphics[width=2.8in]{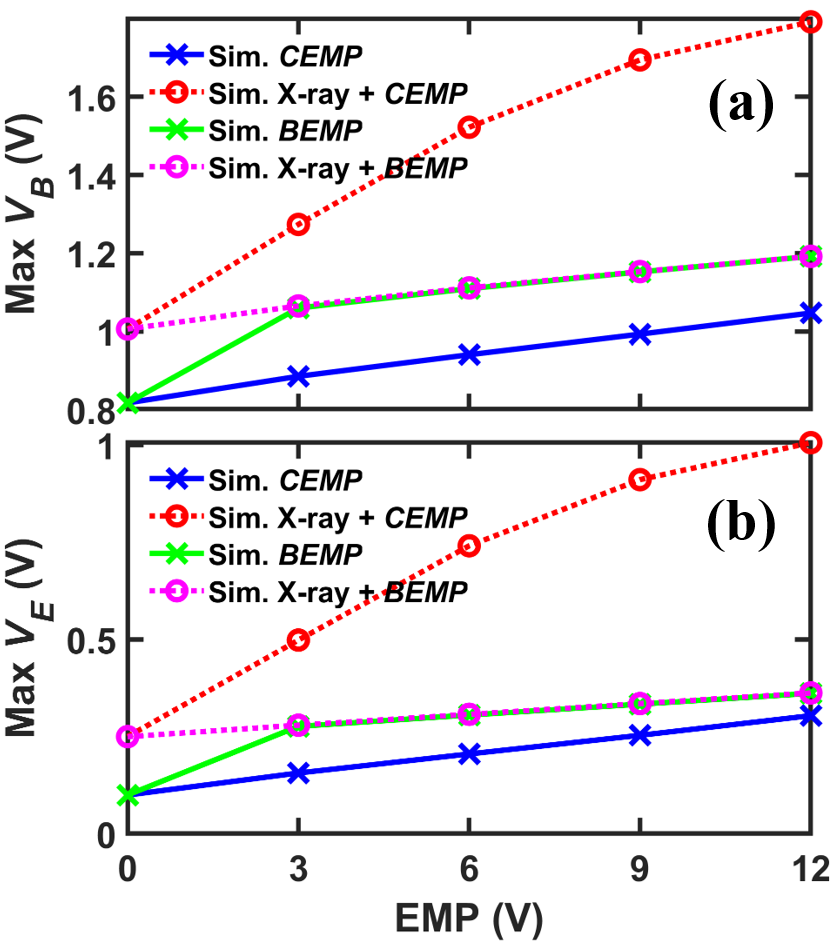}}
  \caption{\hl{Simulation results of the combined effects of pulsed X-ray irradiation and voltage pulse with varying amplitudes: (a) Maximum base voltage response (0-200\,ns); (b) Maximum emitter voltage response (0-200\,ns).}}\label{sim_diff}
\end{figure}

Fig.~\ref{sim_coll} illustrates the simulation results of the combined effects of pulsed X-ray irradiation and CEMP when the peak value of CEMP is set at 9\,V. As shown in Fig.~\ref{sim_coll}(b) and~\ref{sim_coll}(c), when the transistor is operating properly, the base voltage is approximately 0.8\,V and the emitter voltage is approximately 0.1\,V. When exposed to pulsed X-ray irradiation alone, both the base and emitter voltage responses exhibit positive perturbations lasting approximately 1\,µs, with the maximum perturbations of approximately 0.2\,V at the base and 0.15\,V at the emitter. When only CEMP is applied, the voltage responses at both the base and emitter initially exhibit brief positive perturbations followed by brief negative perturbations, with the maximum positive perturbations reaching approximately 0.25\,V. Simultaneous exposure to pulsed X-ray irradiation and CEMP results in positive voltage perturbations at both the base and emitter, with maximum positive perturbations of approximately 1.0\,V at the base and 0.9\,V at the emitter. The black waveform (combined effect minus CEMP-only effect) is much larger than the red waveform (X-ray-only effect), indicating that the combined effect of pulsed X-ray and CEMP exceeds the linear superposition of their individual effects, consistent with the experimental results.

Fig.~\ref{sim_base}  illustrates the simulation results of the combined effects of pulsed X-ray irradiation and BEMP when the peak value of BEMP is set at 9\,V. As shown in Fig.~\ref{sim_base}(b) and~\ref{sim_base}(c), when pulsed X-ray irradiation acts alone, both the base and emitter voltage responses exhibit positive perturbations lasting approximately 1\,µs, with maximum perturbations of approximately 0.2\,V at the base and 0.15\,V at the emitter. When only BEMP is applied, the voltage responses at both the base and emitter exhibit positive perturbations, with maximum positive perturbations reaching approximately 0.3\,V at the base and 0.25\,V at the emitter. Simultaneous exposure to pulsed X-ray irradiation and BEMP results in positive voltage perturbations at both the base and emitter, with maximum positive perturbations of approximately 0.3V at the base and 0.25\,V at the emitter. The black waveform (combined effect minus BEMP-only effect) is close to zero, indicating that the combined effect of pulsed X-ray irradiation and BEMP closely resembles the effect of BEMP acting alone, consistent with the experimental findings.

It should be noted that comparing Fig.~\ref{sim_coll} and~\ref{sim_base} with Fig.~\ref{mea_coll} and~\ref{mea_base}  reveals that the TCAD simulation results do not reproduce the negative perturbations preceding the positive ones observed in the experimental waveforms. This further confirms that the negative pulses in the experiment originated from crosstalk interference in the test system, constituting an artifact superimposed on the true transistor response. Therefore, the subsequent analysis in this study focuses primarily on the maximum positive perturbation values at the base and emitter. Additionally, the positive perturbation waveforms in the TCAD simulation results are sharp peaks, whereas the experimental results show flat-topped waveforms. This discrepancy arises because the 2D transistor model used in the TCAD simulation cannot perfectly match the actual device in terms of dimensions, doping profiles, and carrier lifetimes, leading to a shorter duration of the diffusion component of the photocurrent response in the simulation compared to the experiment.

However, observing the peak pulse response reveals that the trends of the combined effects observed in the TCAD simulation results are entirely consistent with the experimental results: the combined effect of pulsed X-ray and CEMP exceeds linear superposition, while the combined effect with BEMP resembles the BEMP-only effect.

Fig.~\ref{sim_diff} summarizes the simulation results of the maximum positive perturbations within the 0-200\,ns time range for (a) base voltage and (b) emitter voltage under different voltage pulse amplitudes. The results demonstrate that the combined effects observed with voltage pulse of peak values at 3\,V, 6\,V, 9\,V and 12\,V are consistent. It is important to note that the focus of this study is the experimental exploration of the combined effects of transient X-ray and EMP, utilizing 2D simulation for mechanism analysis. Although the simulation waveforms do not perfectly match the experimental waveforms (negative pulses in the experiment are attributed to crosstalk in the complex system, and sharp peaks instead of plateaus in the simulation arise from model approximations), the simulation and experiment show high consistency in the trends of the combined effects as manifested in the pulse response peaks. More importantly, experiments and simulations under multiple different EMP amplitudes consistently verified this trend. Therefore, the current simulations provide valuable insights into the physical factors dominating the combined effects. We will continue to optimize the experimental system and simulation models in future research to achieve more accurate measurements and higher-fidelity simulations.

\begin{figure*}
  \centerline{\includegraphics[width=7.0in]{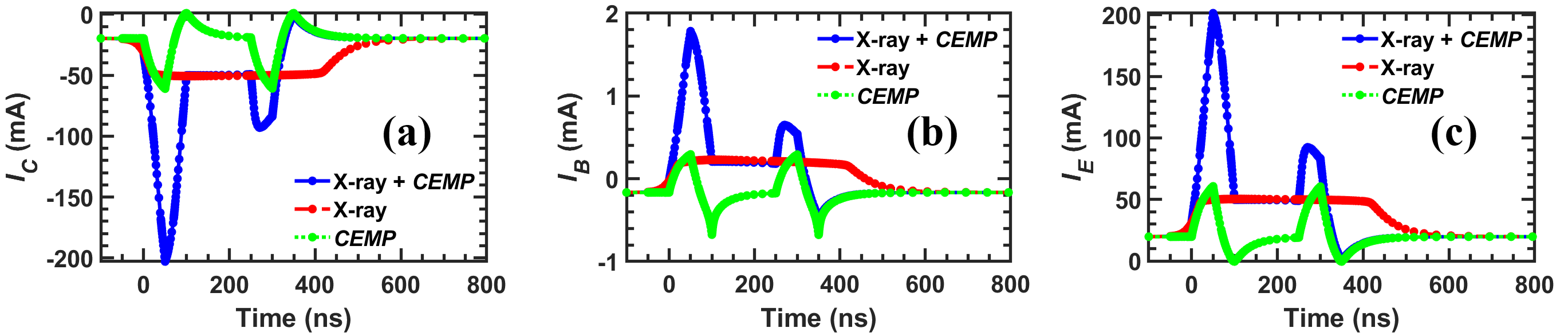}}
  \caption{Simulation results of current response for the combined effect of pulsed X-ray irradiation and CEMP (peak 9\,V): (a) Collector current $I_{C}$; (b) Base current $I_{B}$; (c) Emitter current $I_{E}$.}\label{sim_cI}
\end{figure*}

\begin{figure*}
  \centerline{\includegraphics[width=7.0in]{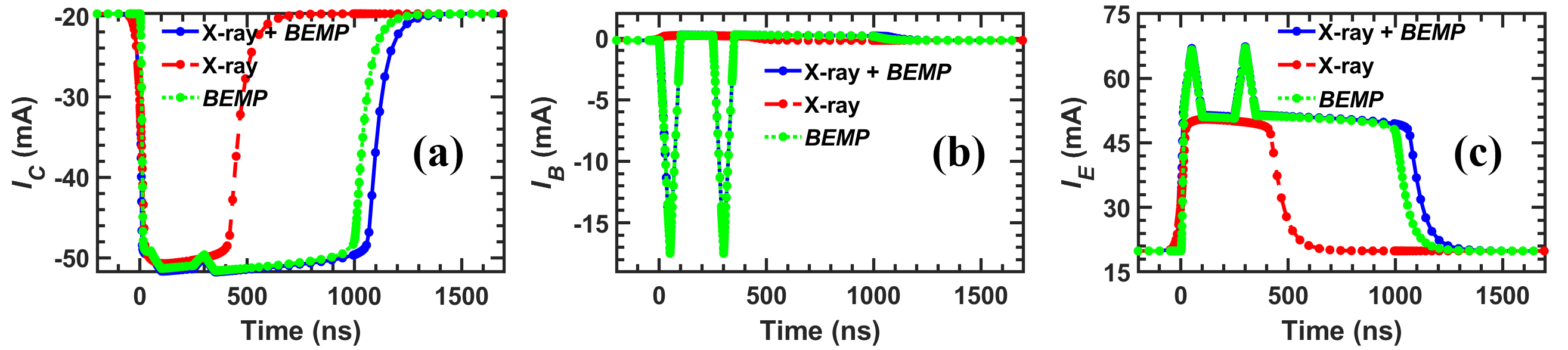}}
  \caption{Simulation results of current response for the combined effect of pulsed X-ray irradiation and BEMP (peak 9\,V): (a) Collector current $I_{C}$; (b) Base current $I_{B}$; (c) Emitter current $I_{E}$.}\label{sim_bI}
\end{figure*}


\subsection{Mechanism Analysis}

Fig.~\ref{sim_cI} shows the simulation results of the current response for the combined effect of pulsed X-ray irradiation and CEMP with the peak amplitude set to 9 V. Fig.~\ref{sim_cI} (a) shows the collector current $I_{C}$, Fig.~\ref{sim_cI} (b) shows the base current $I_{B}$, and Fig.~\ref{sim_cI}(c) shows the emitter current $I_{E}$. It can be seen that under transient X-ray, CEMP alone, or their combined action, the collector current and emitter current initially increase and then gradually recover compared to the normal operating current; whereas the base current first reverses direction (becomes negative) and then gradually recovers. The maximum current response at any electrode is about 200\,mA. Fig.~\ref{sim_bI} shows the simulation results of the current response for the combined effect of pulsed X-ray irradiation and BEMP with the peak amplitude set to 9 V. Fig.~\ref{sim_bI} (a) shows the collector current $I_{C}$, Fig.~\ref{sim_bI}(b) shows the base current $I_{B}$, and Fig.~\ref{sim_cI}(c) shows the emitter current $I_{E}$. It can be seen that under transient X-ray, BEMP alone, or their combined action, the collector current and emitter current initially increase and then recover. Under BEMP alone or combined action, the base current first increases rapidly and then decreases rapidly. The maximum current response at any electrode does not exceed 100\,mA. Reference \cite{wu2023combined} conducted temperature simulations on the pulse burnout effect of the 2N2222A, indicating that a current response of about 2\,A with a duration greater than 400\,ns is required to cause burnout in the 2N2222A. Therefore, the TCAD mechanism analysis in this paper temporarily does not consider the synergy between EMP-induced temperature rise and transient X-ray, but instead focuses on the synergy between EMP-induced electric field changes and transient X-ray. Moreover, as shown in Figs.~\ref{mea_base_diff} and ~\ref{sim_diff}, the trends of the combined effects obtained from the simulations without considering temperature rise already match the experimental trends, indicating that the dominant mechanism for the combined effects studied here is the synergy between EMP-induced electric field changes and transient X-ray.


Existing research indicates that the photocurrent generated in transistors under pulsed X-ray irradiation consists of two components: first, the nanosecond-order drift photocurrent formed by carriers within the depletion region under the influence of the electric field; and second, the diffusion photocurrent formed by carriers generated outside the depletion region that diffuse to the depletion region are collected, constituting the diffusion component of the photocurrent \cite{carr1964transient, wirth1964transient, fjeldly2001modeling, alexander2003transient}.

\begin{figure*}
  \centerline{\includegraphics[width=6.4in]{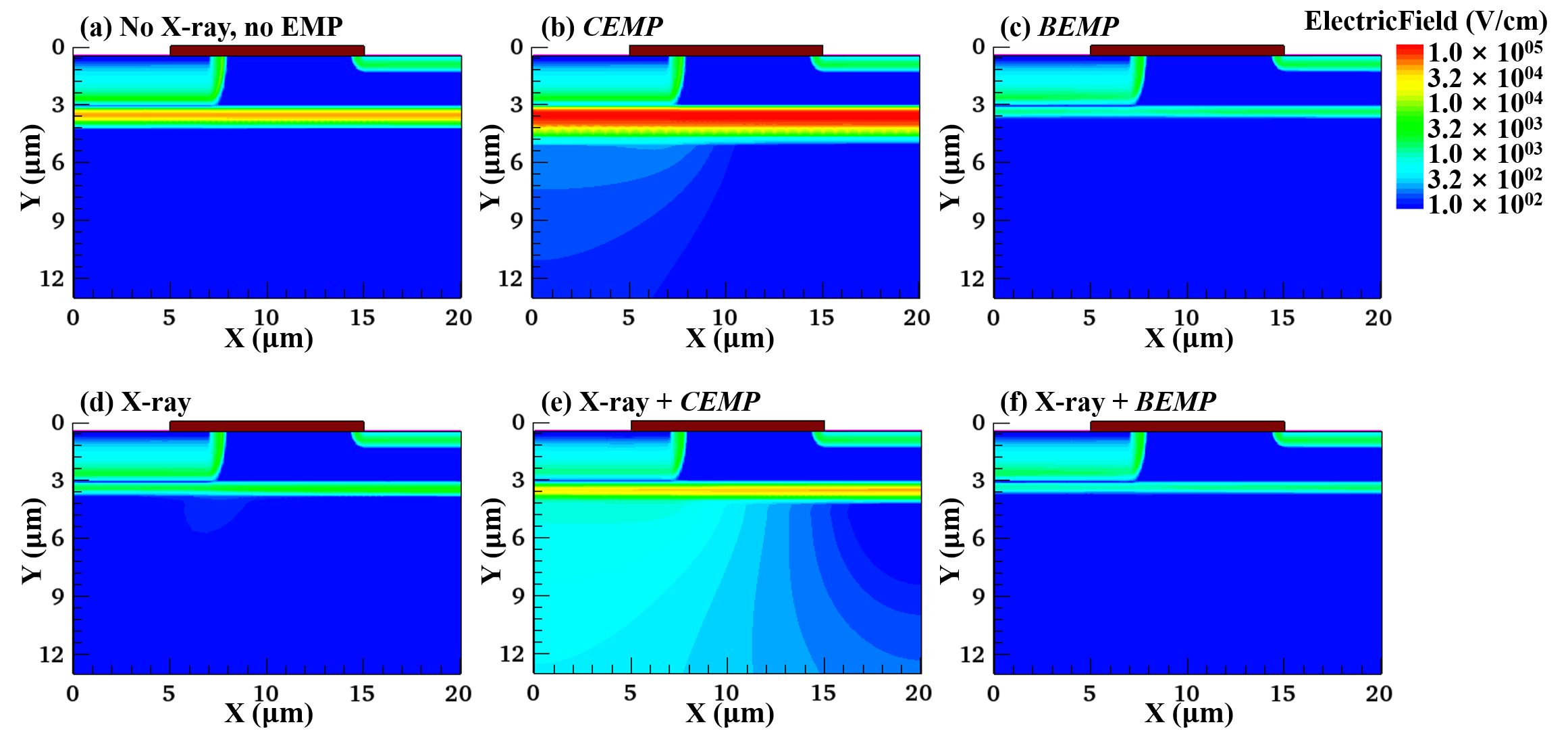}}
  \caption{Electric field strength distribution at t = 50\,ns: (a) Normal operation; (b) CEMP alone; (c) BEMP alone; (d) Pulsed X-ray irradiation alone; (e) Pulsed X-ray irradiation and CEMP simultaneously; (f) Pulsed X-ray irradiation and BEMP simultaneously.}\label{tcad_ef}
\end{figure*}

Fig.~\ref{tcad_ef} shows the electric field distribution at t\,=\,50\,ns. As shown in Fig.~\ref{tcad_ef}(a), the electric field distribution when the transistor operates in forward active mode (emitter junction forward-biased, collector junction reverse-biased). As shown in Fig.~\ref{tcad_ef}(b), when CEMP acts alone, the reverse bias across the collector junction increases significantly, leading to enhanced electric field strength and depletion region width. Conversely, as shown in Fig.~\ref{tcad_ef}(c), when BEMP acts alone, the collector junction becomes forward-biased, resulting in reduced electric field strength and depletion region width. Therefore, when transient X-ray and CEMP act together, the drift photocurrent at the collector junction increases significantly; however, when transient X-ray and BEMP act together, the drift photocurrent is small.

\begin{figure*}
  \centerline{\includegraphics[width=6.4in]{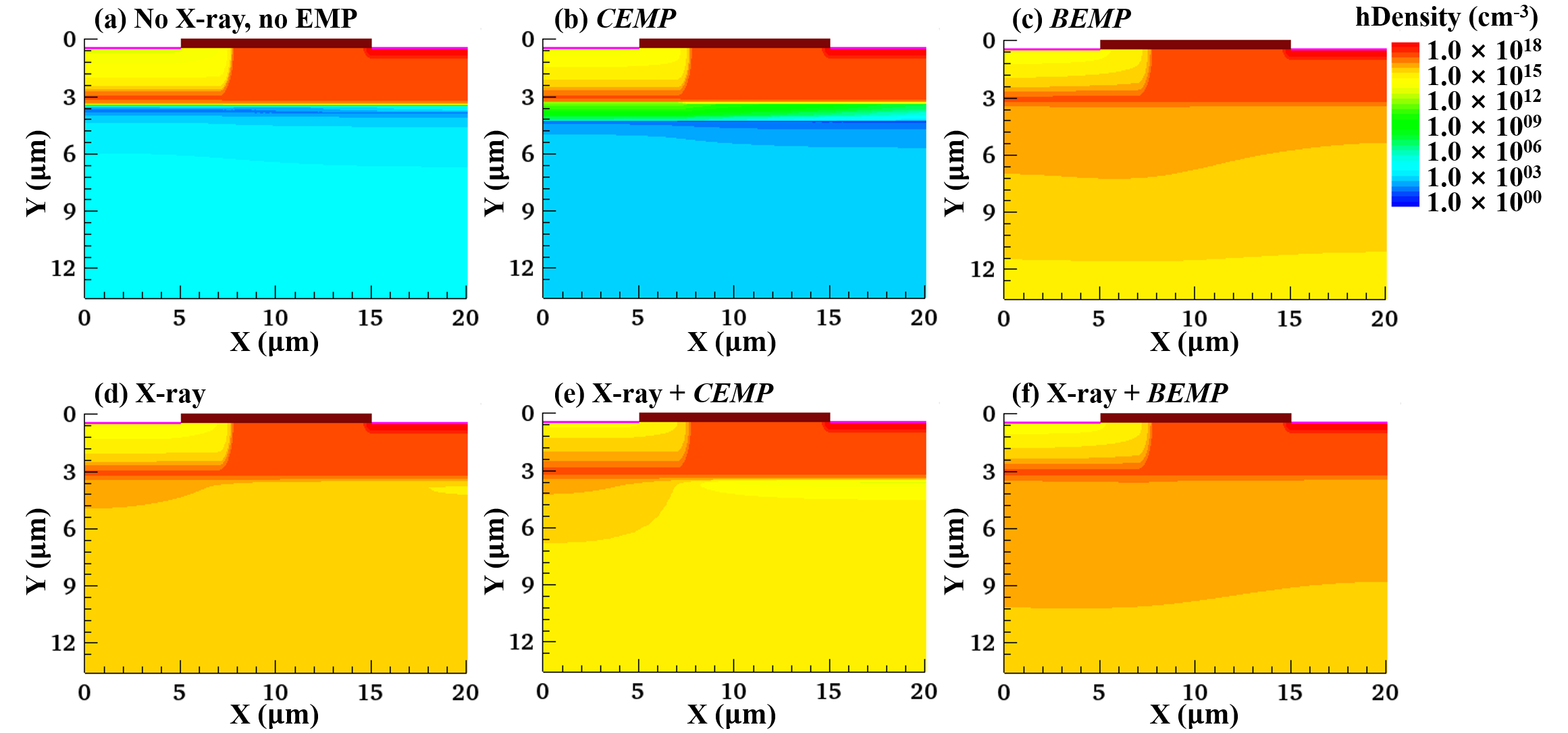}}
  \caption{Hole density distribution at t = 50\,ns: (a) Normal operation; (b) CEMP acted alone; (c) BEMP acted alone; (d) Pulsed X-ray irradiation acted alone; (e) Pulsed X-ray irradiation and CEMP acted simultaneously; (f) Pulsed X-ray irradiation and BEMP acted simultaneously.}\label{tcad_hole}
\end{figure*}

Fig.~\ref{tcad_hole} shows the hole density distribution at t\,=\,50\,ns. As shown in Fig.~\ref{tcad_hole}(b), when CEMP acts alone, the transistor remains in forward active mode, and the hole concentration gradient in the collector region is higher than under normal operating conditions. Conversely, as shown in Fig.~\ref{tcad_hole}(c), when BEMP acts alone, the transistor enters saturation mode, and the hole concentration in the collector region is significantly higher than under normal operating conditions. Therefore, when pulsed X-ray irradiation and CEMP act simultaneously, the diffusion photocurrent near the collector junction increases significantly. In contrast, when pulsed X-ray irradiation and BEMP act simultaneously, the diffusion photocurrent is small.

In summary, when pulsed X-ray irradiation and CEMP act simultaneously, the transistor remains in forward active mode, and both the drift photocurrent at the collector junction and the diffusion photocurrent near the collector junction are significantly enhanced. Therefore, their combined effect exceeds the linear superposition of their individual effects. Whereas, when pulsed X-ray irradiation and BEMP act simultaneously, the transistor enters saturation mode, and both the drift photocurrent at the collector junction and the diffusion photocurrent near the collector junction are small. Therefore, their combined effect is very similar to the effect of BEMP acting alone.

\section{Conclusion}
This paper thoroughly investigates the combined effects of transient ionizing radiation and electromagnetic pulse (EMP) on vertical NPN bipolar transistors through pulsed X-ray irradiation experiments and TCAD simulations. Both experimental and simulation results demonstrate that the combined effect of pulsed X-ray irradiation and \hl{a pulse injected into the collector (CEMP)} exceeds the linear superposition of their individual effects, whereas the combined effect of pulsed X-ray irradiation and \hl{a pulse injected into the base (BEMP)} aligns closely with the effect of BEMP acting alone. 

Further mechanism analysis reveals that when CEMP acts alone, the transistor maintains its forward active mode. This leads to a significant increase in the reverse bias across the collector junction, thereby enhancing the electric field strength and depletion region width. Additionally, the electron concentration gradient in the base and the hole concentration gradient in the collector exceed the levels under normal operating conditions. Conversely, when BEMP acts alone, the transistor transitions into saturation mode, causing the collector junction to become forward-biased. This results in reduced electric field strength and depletion region width, while the electron and hole concentrations in the base and collector significantly exceed the densities under normal operation. Consequently, simultaneous exposure to pulsed X-ray irradiation and CEMP causes a significant increase in both the drift photocurrent at the collector junction and the diffusion photocurrent near the collector junction. In contrast, when pulsed X-ray irradiation and BEMP act simultaneously, these photocurrent components remain small.

This finding applies to all scenarios where transient X-rays and EMP coexist (e.g. inside ICF facilities). These findings provide important guidance for the radiation-hardening design of bipolar circuits in extreme radiation environments.

\section*{Acknowledgment}
The authors would like to thank Shuqin Ren and Zenggang Wu at the Northwest Institute of Nuclear Technology for their invaluable assistance during the experiment. 


\end{document}